\documentclass[twocolumn,showpacs,prl,aps,floatfix]{revtex4-1}

\usepackage{graphicx}
\usepackage{amsmath}
\usepackage[center]{subfigure}

\begin{document}

\title{Theory of cooperation in a micro-organismal snowdrift game}
\author{Zhenyu Wang and Nigel Goldenfeld}
\affiliation{Department of Physics, Center for the Physics of Living
Cells and Institute for Genomic Biology, University of Illinois at Urbana-Champaign, Loomis Laboratory of
Physics, 1110 West Green Street, Urbana, Illinois, 61801}

\begin{abstract}
We present a mean field model for the phase diagram of a community of
micro-organisms, interacting through their metabolism so
that they are, in effect, engaging in a cooperative social game.  We
show that as a function of the concentration of the nutrients glucose
and histidine, the community undergoes a phase transition separating a
state in which one strain is dominant to a state which is characterized
by coexisting populations.  Our results are in good agreement with
recent experimental results, correctly reproducing quantitative trends
and predicting the phase diagram.

\end{abstract}


%
%
%
%
%
%
%
%
%
%
%
%
%

\maketitle

Cooperative phenomena in biology are difficult to treat because of the
complexity and heterogeneity of the interactions, but a qualitatively
successful approach is cooperative game theory---the effort to
encapsulate the complex interactions into parameters describing the
binary outcome of pairwise interactions between
individuals\cite{smit82,haue05,axel06,nowa06,pros07,anta09,trau10}. The
central element in game theory is the payoff matrix, which describes
the score accruing to each member of an interacting pair depending upon
their action in the game. For example, in the Prisoner's Dilemma, the
two players can either \lq\lq cooperate" or \lq\lq defect". Mutual cooperation
yields a reward $R$, whilst if both defect, they receive a punishment
$P$. If one defects and the other cooperates, the defector receives a
temptation $T$ while the cooperator receives the sucker's payoff $S$.
If $T>R>P>S$, then there is a dilemma: a rational player would defect
to receive the highest payoff independent of the state of the other
player, so that if both parties play rationally, each will end up with
the punishment $P$. However, if they had both cooperated, they would
have received the reward $R$.

Two-body interactions are paradoxical in cooperative games, a forceful
indicator of how collective effects can override selfish one-body
behavior. If the the payoff matrix instead obeyed the inequalities
$T>R>S>P$, then the rational strategy is to do the opposite of the
other player. This condition leads to the so-called Snowdrift Game,
which corresponds to a coexistence of players.

Such seemingly abstract games have biological realizations in the
dynamics of microbes and viruses.
%
%
%
In a recent experiment, a game theory payoff matrix was manipulated
by genetically engineering \textit{Saccharomyces cerevisiae} (budding
yeast)\cite{gore09}. Budding yeast's primary carbon intake is a
monosaccharide, such as glucose and fructose. In a
monosaccharide-absent environment, dormant genes are derepressed to
digest alternative nutrients, such as disaccharide maltose and
sucrose\cite{ganc98}. In the experiment, wild-type cooperative strains
have an intact \textit{SUC2} gene, which codes enzyme invertase to
hydrolyze sucrose into glucose and fructose. However, $99\%$ of the
product is released back into the media, giving rise to the possibility
that mutant defectors with the \textit{SUC2} gene knocked out could
make use of the metabolite without having to pay the price of
manufacturing glucose. In order to tune the cost of cooperation and
hence the payoff matrix, the authors engineered cooperators to be
histidine auxotrophs, relying on histidine importation from the media.
Having an intact histidine gene, defectors are not affected. Thus
limitation of histidine concentration in the media coerces the
metabolism of cooperators, increases the cost of cooperation, and thus
affects the payoff matrix. By changing the glucose and histidine
concentration provided with a fixed portion of sucrose, the authors
empirically obtained a transition from the dominance of defectors,
which corresponds to the Prisoner's Dilemma, to the coexistence of both
strains, which is a Snowdrift Game. The ability to manipulate
collective properties of the microbial world by genetic engineering is
impressive, but what is lacking is a predictive understanding of the
direct dependence of cooperator fraction on nutrition concentrations.

The purpose of this paper is to build up a phenomenological model
linking game theory and experimental measurable quantities. We
calculate the population structure, i.e. the fraction for cooperators
and defectors, at different glucose and histidine concentrations, and
reproduce the phase diagram for the transition from dominance of a
single strain to coexistence of both. We use phenomenological game
theory because the collective effects here are highly nonlinear due to
complex metabolism. Our model implies a consistent nonlinearity
responsible for both yeast growth and glucose production.

\begin{figure}[tbp]
\centering
\includegraphics[scale=0.4]{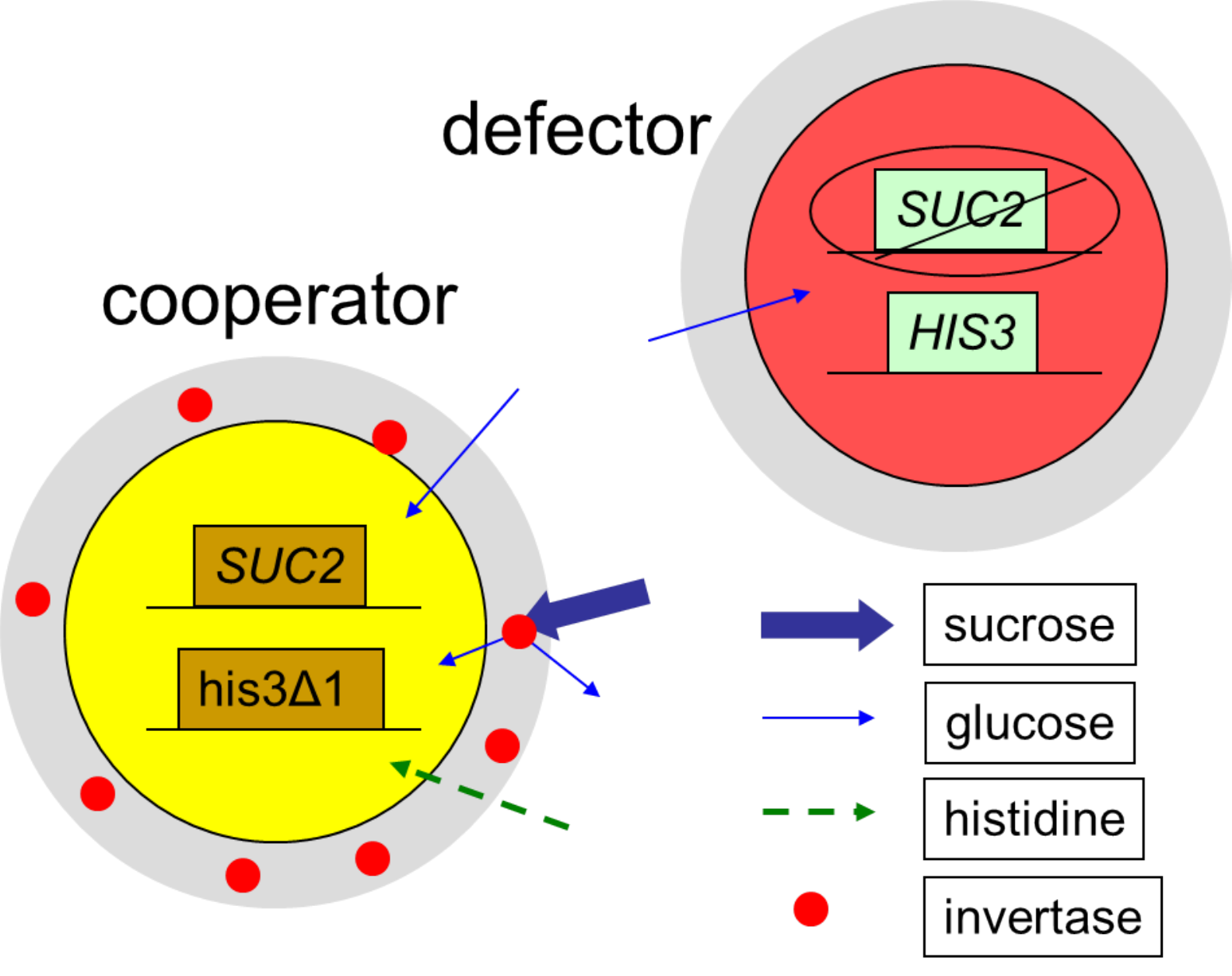}\newline
\caption{(Color online) Schematic of nutrient flows in the experiment of Ref.
\cite{gore09}. Sucrose is hydrolyzed in the periplasmatic space (grey)
of cooperators. The majority of the glucose produced diffuses back to
the media, from which both strains import glucose.}
\label{sequence2}
\end{figure}

The interactions between cooperative and defective strains are
complicated for the following two reasons. First, there are two kinds
of nutritional molecules: sucrose and glucose as sketched in Fig.~(\ref{sequence2}).
Sucrose is easy to handle because it has a single source and single mode
of consumption, originating from the media and being consumed only by
cooperators. However, glucose has two sources: the initial amount
added into the media, and the local increment from sucrose
decomposition by cooperators. The actual glucose concentration
surrounding yeast cells depends on the cooperators' metabolism and
concentration, whose relation is unknown. Second, in sucrose
hydrolysis, cooperators experience a cost to synthesize invertase,
but at the same time gain in generating glucose for
themselves. The balance between the cost and benefit is
subtle and hard to handle. In order to circumvent these two obstacles,
we model a simple situation where the two strains are at the same
nutrition level. This should be applicable to the experimental
situation because cooperative strains ultimately live on the
monosaccharide glucose no matter if it is absorbed from the surrounding
media or decomposed from sucrose. In this way, our system can be
simplified as a coexistence problem of two strains living on the same
nutrition glucose.

Next we use game theory to identify the conditions for coexistence. The
key is to construct a payoff matrix with experimental data. Here, the
two strains are engaging in a cooperative game: if the payoff for
defectors exceeds that of cooperators, defectors will dominate; if the
payoff for cooperators exceeds that of defectors, cooperators will
dominate. Therefore only when the payoffs for both parties are equal,
will coexistence be achieved. The payoff for players is the mean
fitness for strains, which is measured as the growth rate. Thus, our
next task is to construct the dependency of growth rates on
experimental observable quantities. We do this below using a mean
field theory, modeled after the way in which cooperative interactions
leading to ferromagnetism are described by an effective local field that
adds to the externally applied magnetic field (see, e.g. Ref.
{\cite{GOLD92}}).

The first input is the nonlinear dependency of growth rate $b$
(hr$^{-1}$) on glucose concentration $g$ ($\%$) according to the
experiment\cite{gore09}:
\begin{equation}
b = \gamma_1g^\alpha,
\label{b0}
\end{equation}
\noindent where $\gamma_1 = 0.44$, $\alpha = 0.15$ and $g$ is
$0.001\sim0.03\%$.
In Eq. (\ref{b0}), the growth rate $b$ varies nonlinearly with glucose
concentration $g$. The nonlinear power $\alpha$ is unusual, and
reflects cellular constraints, such as the nonlinear performance of
hexose transporters and catabolic pathway enzymes. We cannot use first
principles system biology to justify the nonlinear $\alpha$, because
the basic metabolic networks etc are not well enough understood.
Instead, we make a very simplified assumption: we interpret the
nonlinearity as primarily reflecting aspects of the efficiency of
hexose transporters across the cell membrane.  Hence Eq.~(\ref{b0}) implies
that translocation flux rate through the membrane is proportional to
the concentration raised to a nonlinear power $\alpha$. Note that in
principle, such translocation processes are influenced by the
metabolism of the cells, but for now we regard that as negligible.

Second, we include the presence of cooperators. Now, there are two
sources of glucose. Besides the initial glucose added into the media,
cooperators also produce glucose from sucrose decomposition. At the
mean field level, every cooperator manufactures glucose at about the
same rate. We assume that this rate does not have a significant
dependence on the metabolism of cells; because the amount of invertase
in each cell is not influenced by the metabolism, we assume that the
performance of invertase is also not significantly influenced by the
metabolism. Since the sucrose concentration is kept the same throughout
the experiment, there is no need for us to explore the detailed form of
such a production rate. The total glucose produced inside all the
cooperator cells is thus proportional to the cooperator fraction $f$.
Eq.~(\ref{b0}) implies that the glucose imported into the cell scales
as $g^{\alpha}$ due to the cellular constraints on the molecular
translocation process. The same translocation passage limits the
glucose output from cooperators, as evidenced by the report that the
diffusion coefficient through the cell wall is anomalously small,
estimated to be $1/20$ of that in water\cite{gore09}.
Hence the flux of glucose released is
proportional to the glucose produced inside the cells raised to the
power $\alpha$. Since the glucose manufactured inside the cells is
proportional to the cooperator fraction $f$, the glucose contribution
from cooperators is proportional to $f^\alpha$ with some coefficient of
proportionality.  We denote the coefficient as $\gamma$. As we note in
the discussion about Eq. (\ref{b0}), the translocation process is
affected by the metabolism of the cells. The coefficient $\gamma$, in
this way, represents a general discount factor due to metabolism, which
is a combined effect of the artificial discount in histidine limitation
and the natural cost in cooperation. Hence, we obtain the growth rate
for defectors
\begin{equation}
b_d = \gamma_1(g + \gamma f^\alpha)^\alpha,
\label{bd}
\end{equation}
\noindent where $\gamma$ is a general discount factor that varies with
histidine concentration, reflecting the artificial discount in
histidine limitation and the natural cost of cooperation.


Third, we analyze the situation for cooperators. Compared with
defectors, when they import glucose from the media, the translocation
process is influenced by the metabolism as we learn from
Eq.~(\ref{b0}). Such a discount, representing a combined effect of the
artificial discount in histidine limitation and the natural cost in
cooperation, is represented by the same $\gamma$ as in Eq. (2), because
the same cellular processes are involved.  Thus we obtain
\begin{equation}
b_c = \gamma\gamma_1(g + \gamma f^\alpha)^\alpha,
\end{equation}
\noindent where $b_c$ is the growth rate for cooperators. Last, we
recall that there is a small amount of glucose that cooperators reserve
for themselves. This amount is determined by the sucrose concentration
and the cell's metabolism and transport processes, which are mediated
by the histidine concentration. Since the sucrose concentration is
always $5\%$ during the experiment, we denote the benefit for a single
cooperator cell by $\zeta$, a single-variable function of
histidine concentration only. Including this benefit for cooperation,
we finally obtain
\begin{equation}
b_c = \gamma\gamma_1(g + \gamma f^\alpha)^\alpha + \zeta.
\label{bc}
\end{equation}

Eqs. (\ref{bd}) and (\ref{bc}) compose the central part of our model,
including the contribution of cooperators to the increase in
glucose concentration by the term $\gamma f^\alpha$. This model
balances the cost $\gamma$ for cooperators with the benefit $\zeta$, both
depending only on histidine concentrations. Note
that as the cooperator fraction $f$ increases, more glucose is trapped
in cooperators, but the amount per cooperator does not change.  The
positivity of $\zeta$ is essential for the survival of cooperators,
which makes it possible for the two engineered strains to engage in a
snowdrift game.

In our model of cooperation, we have input three non-trivial arguments:
(i) The two $\alpha$'s in Eqs. (\ref{bd}) and (\ref{bc}) are the same,
representing the same translocation passage limitation on the glucose flux
both into and out of yeast cells;
(ii) The two
$\gamma$'s in Eq. (\ref{bc}) are the same, implying the same discount
in yeasts' growth and sucrose decomposition by cost of cooperation
mediated by histidine limitation; (iii) $\zeta$ is a single-variable function of
histidine concentration, representing that cooperators are compensated
for production of glucose.

Our arguments above motivated points (i)-(iii) assuming that it is
primarily the phenomenology of transport of glucose through the cell
membrane which is the growth-rate determining factor. However, in
principle, other metabolic effects can be present. To test whether or
not our assumptions are self-consistent and represent a good
representation of the data, we compare the predictions of our equations
with the data.

Ideally, we would like to be able to calculate the cooperator fraction
as a function of glucose and histidine concentrations (Fig. 3 (b) of
Ref. \cite{gore09}) from theory, but this would require a detailed
description of the metabolism and growth dynamics of the organisms to
obtain the parameters. As an alternative approach, we input
experimental data to our equations and verify the consistency of our
modeling by checking the standard deviations for different sets of
data. Based on our reasoning from game theory that the growth rates for
cooperators and defectors are the same at equilibrium, the measured
growth rates of cocultures as a function of glucose and histidine
concentrations (Fig. 3(c) of Ref. \cite{gore09}) should be valid for
either strain. Interpreting them as the growth rates for defectors, we
can import the data in Fig. 3(b-c) of Ref. \cite{gore09} for various
glucose and histidine concentrations into Eq. (\ref{bd}) and calculate
the discount $\gamma$. According to our argument (i), we predict that
$\gamma$ should be the same at the same histidine concentration but
different glucose concentrations; this is supported by the standard
deviations shown in Table \ref{tab:gamma}. We neglect the data for very
small cooperator fractions, especially for the extinction of
cooperators, such as those when histidine concentration is as low as
$0.005$, since they will either generate large deviation with very
small bias in measurement or cause the cooperation term $\gamma
f^\alpha$ to vanish. Averaging among different glucose concentrations,
we can see that the discount $\gamma$ gets smaller when histidine is
more dilute.
The first two $\sigma_\gamma$ are calculated with six data points where
glucose concentration ($\%$) ranges from $0.001$ to $0.03$. The latter
two are smaller than the first two since fewer data are averaged. The
smallness of the standard deviations has not been hard-wired into our
model, and substantiates our assumption (i) because otherwise they
might be orders of magnitude larger, as we will illustrate as follows.
We show in Table \ref{tab:gamma1} the average of $\gamma$ and its
corresponding standard deviation $\sigma_\gamma$ if the increment of
glucose concentration varied not with the same power $\alpha$ as we
have assumed in our model, but
linearly with cooperator fraction,
as we might have initially guessed,
\begin{equation}
b_d = \gamma_1(g + \gamma f)^\alpha,
\label{bd1}
\end{equation}
or even quadratically
\begin{equation}
b_d = \gamma_1(g + \gamma f^2)^\alpha.
\label{bd2}
\end{equation}
The standard deviations $\sigma_\gamma$ in Table \ref{tab:gamma1} are
at least two orders of magnitude larger than those in Table
\ref{tab:gamma}, and are even higher for the fit to Eq. (\ref{bd2}).
The comparison among these tables demonstrates that the standard
deviation is a good test of our assumption, and hence justifies the
self-consistency of our theory.

\begin{table}[htbp]
\caption{Cost $\gamma$ for cooperators at various histidine concentrations.}
\label{tab:gamma}\centering
\begin{tabular}{ccccc}
\hline\hline
[his]/(20 $\mu$g ml$^{-1}$) & 1 & 0.2 & 0.05 & 0.02 \\[0.5ex]
$\gamma$ & 0.19 & 0.14 & 0.061 & 0.027 \\[0.5ex]
standard deviation $\sigma_\gamma$ & 0.02 & 0.02 & 0.006 & 0.006 \\[0.5ex] \hline
\end{tabular}%
\end{table}

\begin{table}[htbp]
\caption{Large standard deviation $\sigma_\gamma$ to fit Eq. (\ref{bd1})
in violation of assumption (i).}
\label{tab:gamma1}\centering
\begin{tabular}{ccccc}
\hline\hline
[his]/(20 $\mu$g ml$^{-1}$) & 1 & 0.2 & 0.05 & 0.02 \\[0.5ex]
$\gamma$ & 1.8 & 12 & 12 & 8 \\[0.5ex]
standard deviation $\sigma_\gamma$ & 1.5 & 9 & 7 & 4 \\[0.5ex] \hline
\end{tabular}%
\end{table}

\begin{table}[htbp]
\caption{Benefit $\zeta$ for cooperators at various histidine concentrations.}
\label{tab:zeta}\centering
\begin{tabular}{ccccc}
\hline\hline
[his]/(20 $\mu$g ml$^{-1}$) & 1 & 0.2 & 0.05 & 0.02 \\[0.5ex]
$\zeta$ & 0.269 & 0.260 & 0.241 & 0.222 \\[0.5ex]
standard deviation $\sigma_\zeta$ & 0.003 & 0.004 & 0.007 & 0.02 \\[0.5ex] \hline
\end{tabular}%
\end{table}

\begin{figure}[htbp]
\centering
\includegraphics[scale=0.495]{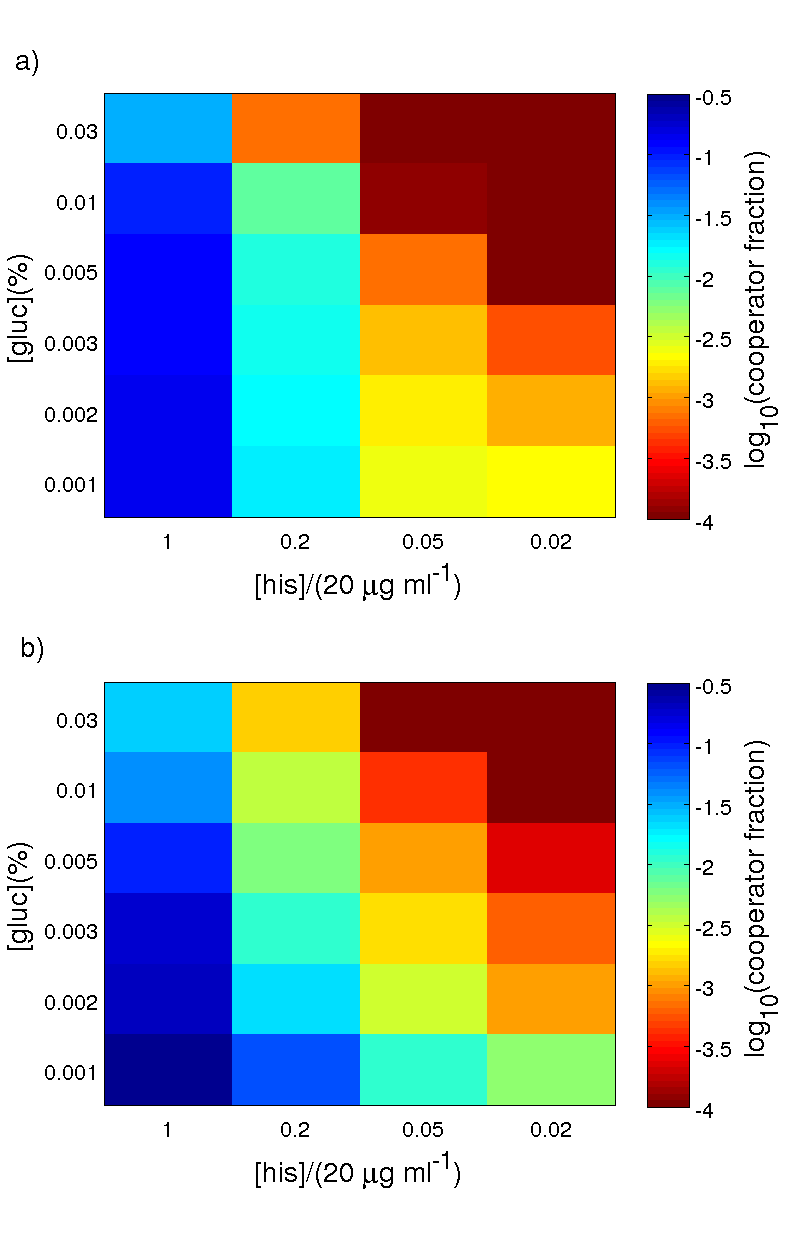}\newline
\caption{(Color online) (a) Theoretical result for cooperator fraction
at various glucose and histidine concentrations. (b) Corresponding
experimental result for cooperator fraction at various glucose and
histidine concentrations.}
\label{pct}
\end{figure}

Next, we interpret the data in Fig. 3(c) of Ref. \cite{gore09} as
growth rates for cooperators and plug in the values of $\gamma$ shown
in Table \ref{tab:gamma} into Eq. (\ref{bc}). Our arguments (ii) and
(iii) predict that $\zeta$ depends only on histidine concentration,
which is consistent with the standard deviation for $\zeta$ in Table
\ref{tab:zeta}. The benefit for cooperators diminishes with the
limitation in histidine. The latter two $\sigma_\zeta$ are bigger than
the previous two since we extend the data for those not incorporated in
the calculation of $\gamma$ in Table \ref{tab:gamma}. Overall,
however, these consistency checks are successful, a result that we
emphasize is not \lq\lq built-in" to our theory.

With the cost $\gamma$ and gain $\zeta$ in hand, we can now
predict the cooperator fraction at equilibrium. Setting $b_d = b_c$ in
Eq. (\ref{bd}) and (\ref{bc}), we plot the predicted cooperator
fraction in Fig. (\ref{pct}a). As a comparison, we replot the
corresponding data from experiment\cite{gore09} in Fig. (\ref{pct}b).
The similarity between the theoretical calculation and experimental
measurement is striking and supports our model.

We have proposed a phenomenological model for wild-type cooperative and
mutant defective strains in a mixed media of glucose and sucrose. We
circumvented the obstacle of modeling sucrose decomposition, which
increases glucose concentration, incurs a cost as invertase syntheses
for cooperators, and rewards them with a small fraction of the glucose
produced, by attributing cost and benefit for cooperation to growth
rates. Then we determined the dependency of growth rates for defectors
and cooperators on experimental quantities such as glucose and
histidine concentrations. Despite our approximations, such as averaging
over different glucose concentrations, the resulting calculation of
cooperator fraction at equilibrium is consistent with experimental
observations.  So what did we actually predict?  By requiring that
$b_d=b_c$, we thus found, {\it in a non-circular way}, the condition
for the phase boundary separating the Prisoner's Dilemma phase from the
Snowdrift phase of the system. Our mean field arguments also predict
the trend, i.e. the sign of $\partial f/\partial g$ for fixed histidine
concentration. These methods could be useful in the design of future
experiments to manipulate collective properties of micro-organism
communities.

We thank Jeff Gore for sharing with us his experimental data, and
Kirill Korolev for helpful comments on the manuscript.  This
work was partly supported by National Science Foundation grant
number NSF-EF-0526747.

\bibliography{snowdrift_bib}

\begin{thebibliography}{10}%
\makeatletter
\providecommand \@ifxundefined [1]{%
 \ifx #1\undefined \expandafter \@firstoftwo
 \else \expandafter \@secondoftwo
\fi
}%
\providecommand \@ifnum [1]{%
 \ifnum #1\expandafter \@firstoftwo
 \else \expandafter \@secondoftwo
\fi
}%
\providecommand \enquote [1]{``#1''}%
\providecommand \bibnamefont  [1]{#1}%
\providecommand \bibfnamefont [1]{#1}%
\providecommand \citenamefont [1]{#1}%
\providecommand\href[0]{\@sanitize\@href}%
\providecommand\@href[1]{\endgroup\@@startlink{#1}\endgroup\@@href}%
\providecommand\@@href[1]{#1\@@endlink}%
\providecommand \@sanitize [0]{\begingroup\catcode`\&12\catcode`\#12\relax}%
\@ifxundefined \pdfoutput {\@firstoftwo}{%
 \@ifnum{\z@=\pdfoutput}{\@firstoftwo}{\@secondoftwo}%
}{%
 \providecommand\@@startlink[1]{\leavevmode\special{html:<a href="#1">}}%
 \providecommand\@@endlink[0]{\special{html:</a>}}%
}{%
 \providecommand\@@startlink[1]{%
  \leavevmode
  \pdfstartlink
   attr{/Border[0 0 1 ]/H/I/C[0 1 1]}%
   user{/Subtype/Link/A<</Type/Action/S/URI/URI(#1)>>}%
  \relax
 }%
 \providecommand\@@endlink[0]{\pdfendlink}%
}%
\providecommand \url  [0]{\begingroup\@sanitize \@url }%
\providecommand \@url [1]{\endgroup\@href {#1}{\urlprefix}}%
\providecommand \urlprefix [0]{URL }%
\providecommand \Eprint[0]{\href }%
\@ifxundefined \urlstyle {%
  \providecommand \doi [1]{doi:\discretionary{}{}{}#1}%
}{%
  \providecommand \doi [0]{doi:\discretionary{}{}{}\begingroup
  \urlstyle{rm}\Url }%
}%
\providecommand \doibase [0]{http://dx.doi.org/}%
\providecommand \Doi[1]{\href{\doibase#1}}%
\providecommand \bibAnnote [3]{%
  \BibitemShut{#1}%
  \begin{quotation}\noindent
    \textsc{Key:}\ #2\\\textsc{Annotation:}\ #3%
  \end{quotation}%
}%
\providecommand \bibAnnoteFile [2]{%
  \IfFileExists{#2}{\bibAnnote {#1} {#2} {\input{#2}}}{}%
}%
\providecommand \typeout [0]{\immediate \write \m@ne }%
\providecommand \selectlanguage [0]{\@gobble}%
\providecommand \bibinfo [0]{\@secondoftwo}%
\providecommand \bibfield [0]{\@secondoftwo}%
\providecommand \translation [1]{[#1]}%
\providecommand \BibitemOpen[0]{}%
\providecommand \bibitemStop [0]{}%
\providecommand \bibitemNoStop [0]{.\EOS\space}%
\providecommand \EOS [0]{\spacefactor3000\relax}%
\providecommand \BibitemShut [1]{\csname bibitem#1\endcsname}%
\bibitem{smit82}%
  \BibitemOpen
  \bibfield{author}{%
  \bibinfo {author} {\bibfnamefont{J.}~\bibnamefont{Smith}},\ }%
  \emph{\bibinfo {title} {{Evolution and the Theory of Games}}}\ (\bibinfo
  {publisher} {Cambridge Univ Press},\ \bibinfo {year} {1982})%
  \bibAnnoteFile{NoStop}{smit82}%
\bibitem{haue05}%
  \BibitemOpen
  \bibfield{author}{%
  \bibinfo {author} {\bibfnamefont{C.}~\bibnamefont{Hauert}}\ and\ \bibinfo
  {author} {\bibfnamefont{G.}~\bibnamefont{Szab{\'o}}},\ }%
  \bibfield{journal}{%
  \bibinfo {journal} {Am. J. Phys.}\ }%
  \textbf{\bibinfo {volume} {73}},\ \bibinfo {pages} {405} (\bibinfo {year}
  {2005})%
  \bibAnnoteFile{NoStop}{haue05}%
\bibitem{axel06}%
  \BibitemOpen
  \bibfield{author}{%
  \bibinfo {author} {\bibfnamefont{R.}~\bibnamefont{Axelrod}}, \bibinfo
  {author} {\bibfnamefont{D.}~\bibnamefont{Axelrod}},\ and\ \bibinfo {author}
  {\bibfnamefont{K.}~\bibnamefont{Pienta}},\ }%
  \bibfield{journal}{%
  \bibinfo {journal} {Proc. Natl. Acad. Sci. USA}\ }%
  \textbf{\bibinfo {volume} {103}},\ \bibinfo {pages} {13474} (\bibinfo {year}
  {2006})%
  \bibAnnoteFile{NoStop}{axel06}%
\bibitem{nowa06}%
  \BibitemOpen
  \bibfield{author}{%
  \bibinfo {author} {\bibfnamefont{M.}~\bibnamefont{Nowak}},\ }%
  \bibfield{journal}{%
  \bibinfo {journal} {Science}\ }%
  \textbf{\bibinfo {volume} {314}},\ \bibinfo {pages} {1560} (\bibinfo {year}
  {2006})%
  \bibAnnoteFile{NoStop}{nowa06}%
\bibitem{pros07}%
  \BibitemOpen
  \bibfield{author}{%
  \bibinfo {author} {\bibfnamefont{J.}~\bibnamefont{Prosser}} \emph{et~al.},\
  }%
  \bibfield{journal}{%
  \bibinfo {journal} {Nature Rev. Microbiol.}\ }%
  \textbf{\bibinfo {volume} {5}},\ \bibinfo {pages} {384} (\bibinfo {year}
  {2007})%
  \bibAnnoteFile{NoStop}{pros07}%
\bibitem{anta09}%
  \BibitemOpen
  \bibfield{author}{%
  \bibinfo {author} {\bibfnamefont{T.}~\bibnamefont{Antal}}, \bibinfo {author}
  {\bibfnamefont{H.}~\bibnamefont{Ohtsuki}}, \bibinfo {author}
  {\bibfnamefont{J.}~\bibnamefont{Wakeley}}, \bibinfo {author}
  {\bibfnamefont{P.}~\bibnamefont{Taylor}},\ and\ \bibinfo {author}
  {\bibfnamefont{M.}~\bibnamefont{Nowak}},\ }%
  \bibfield{journal}{%
  \bibinfo {journal} {Proc. Natl. Acad. Sci. USA}\ }%
  \textbf{\bibinfo {volume} {106}},\ \bibinfo {pages} {8597} (\bibinfo {year}
  {2009})%
  \bibAnnoteFile{NoStop}{anta09}%
\bibitem{trau10}%
  \BibitemOpen
  \bibfield{author}{%
  \bibinfo {author} {\bibfnamefont{A.}~\bibnamefont{Traulsen}} \emph{et~al.},\
  }%
  \bibfield{journal}{%
  \bibinfo {journal} {Proc. Natl. Acad. Sci. USA}\ }%
  \textbf{\bibinfo {volume} {107}},\ \bibinfo {pages} {2962} (\bibinfo {year}
  {2010})%
  \bibAnnoteFile{NoStop}{trau10}%
\bibitem{gore09}%
  \BibitemOpen
  \bibfield{author}{%
  \bibinfo {author} {\bibfnamefont{J.}~\bibnamefont{Gore}}, \bibinfo {author}
  {\bibfnamefont{H.}~\bibnamefont{Youk}},\ and\ \bibinfo {author}
  {\bibfnamefont{A.}~\bibnamefont{Van~Oudenaarden}},\ }%
  \bibfield{journal}{%
  \bibinfo {journal} {Nature}\ }%
  \textbf{\bibinfo {volume} {459}},\ \bibinfo {pages} {253} (\bibinfo {year}
  {2009})%
  \bibAnnoteFile{NoStop}{gore09}%
\bibitem{ganc98}%
  \BibitemOpen
  \bibfield{author}{%
  \bibinfo {author} {\bibfnamefont{J.}~\bibnamefont{Gancedo}},\ }%
  \bibfield{journal}{%
  \bibinfo {journal} {Micro. and Mol. Biol. Rev.}\ }%
  \textbf{\bibinfo {volume} {62}},\ \bibinfo {pages} {334} (\bibinfo {year}
  {1998})%
  \bibAnnoteFile{NoStop}{ganc98}%
\bibitem{GOLD92}%
  \BibitemOpen
  \bibfield{author}{%
  \bibinfo {author} {\bibfnamefont{N.}~\bibnamefont{Goldenfeld}},\ }%
  \emph{\bibinfo {title} {Lectures on phase transitions and the renormalization
  group}}\ (\bibinfo {publisher} {Addison-Wesley, Reading, MA},\ \bibinfo
  {year} {1992})%
  \bibAnnoteFile{NoStop}{GOLD92}%
\end{thebibliography}%

\end{document}